# Three-Dimensional Radiotherapy Dose Prediction on Head and Neck Cancer Patients with a Hierarchically Densely Connected U-net Deep Learning Architecture

Dan Nguyen, Xun Jia, David Sher, Mu-Han Lin, Zohaib Iqbal, Hui Liu, Steve Jiang

Medical Artificial Intelligence and Automation Laboratory, Department of Radiation Oncology, University of Texas Southwestern Medical Center, Dallas, TX, 75390, USA

Email: Dan.Nguyen@UTSouthwestern.edu

**ABSTRACT**

The treatment planning process for patients with head and neck (H&N) cancer is regarded as one of the most complicated due to large target volume, multiple prescription dose levels, and many radiation-sensitive critical structures near the target. Treatment planning for this site requires a high level of human expertise and a tremendous amount of effort to produce personalized high quality plans, taking as long as a week, which deteriorates the chances of tumor control and patient survival. To solve this problem, we propose to investigate a deep learning-based dose prediction model, Hierarchically Densely Connected U-net, based on two highly popular network architectures: U-net and DenseNet. We find that this new architecture is able to accurately and efficiently predict the dose distribution, outperforming the other two models, the Standard U-net and DenseNet, in homogeneity, dose conformity, and dose coverage on the test data. Averaging across all organs at risk, our proposed model is capable of predicting the organ-at-risk max dose within 6.3% and mean dose within 5.1% of the prescription dose on the test data. The other models, the Standard U-net and DenseNet, performed worse, having an averaged organ-at-risk max dose prediction error of 8.2% and 9.3%, respectively, and averaged mean dose prediction error of 6.4% and 6.8%, respectively. In addition, our proposed model used 12 times less trainable parameters than the Standard U-net, and predicted the patient dose 4 times faster than DenseNet.

## I. INTRODUCTION

Patients with head and neck (H&N) cancer undergoing radiotherapy have typically been treated with intensity modulated radiation therapy (IMRT)[1-7] and volume modulated arc therapy (VMAT)[8-15], which has significantly reduced toxicity[16-18] and improved quality of life[19,20], as compared to more conventional methods such as 3D conformal radiotherapy. However, treatment planning for this site is regarded as one of the most complicated due to several aspects, including large planning target volume (PTV) size[21], multiple prescription dose levels that are simultaneously integrated boosted[22,23], and many radiation-sensitive organs-at-risk (OAR) that are in close proximity to the PTV[24-27]. Consequently, treatment planning for this site requires a tremendous level of human expertise and effort to produce personalized high quality plans.

In the typical current treatment planning workflow, a treatment planner solves an inverse optimization problem[28], where they adjust a set of hyper-parameters and weightings to control the tradeoffs between clinical objectives. Since the physician preferred plan is largely unknown, the planner meticulously tunes these parameters in a trial-and-error fashion in an attempt to reach an appropriate solution. Many rounds of consultation between the planner and physician occur regarding the plan quality and tradeoffs are discussed. Ultimately, this trial-and-error process in parameter tuning results in hours for a plan to be generated[29-31], and the iterations of consultation between the physician and planner may extend the treatment planning time up to one week. For aggressive H&N tumors, where tumor volume can double in approximately 30 days, which account for 50% of patients[32], an extended planning time can greatly decrease local tumor control and patient survival[33-36].

In recent years, the field of artificial intelligence (AI) and deep learning has made tremendous progress, particularly in the field of computer vision and decision making. In 2015, Ronneberger et al. proposed a deep learning architecture for semantic segmentation, known as U-net[37]. This neural network architecture, a type of convolutional neural network (CNN)[38] that falls under the class fully convolutional networks (FCN)[39], was capable of incorporating both local and global features to make a pixel-wise prediction. These predictions are commonly done slice-by-slice in 2D. For dose prediction, this 2D-based prediction can inherently cause some errors, particularly in slices at the superior and inferior edges of the PTV, thus motivating us to move towards 3D volumetric deep learning models. However, when creating a 3D variant of U-net, the computational expense grows with the dimensionality. Tradeoffs have to be made with the 3D version, such as less filters per convolution or max pooling layers. Attempts to combat this for 3D architectures focused on modifying portions of the architecture to be more efficient at propagating information, such as having a ResNet flavor of including skip connections during each block[40,41]. With the currently available GPU technologies and memory, the network's performance is sacrificed.

A publication in 2017 by Huang et al. proposed a Densely Connected Convolutional Neural Network, also known as DenseNet [42]. The publication proposed the novel idea of densely connecting its convolutional maps together to promote feature propagation and reuse, reduce the vanishing gradient issue, and decrease the number of trainable parameters needed. While the term "densely connected" was historically used to described fully connected neural network layers, this publication by Huang et al. had adopted this terminology to describe how his convolutional layers were connected. While requiring more memory to use, the authors showed that the DenseNet was capable of achieving a better performance while having far less parameters in the neural network. For example, they were able to have comparable accuracy with ResNet, which had 10 million parameters, using their DenseNet, which had 0.8M parameters. This indicates that DenseNet is far more efficient in feature calculation than existing network architectures. For its contribution to the AI community, the DenseNet publication was awarded for the CVPR 2017 best publication. However, it is recognized that DenseNet, while efficient in parameter usage, actually utilizes considerably more GPU RAM, rendering a 3D U-net with fully densely connected convolutional connections infeasible for today's current GPU technologies.

Motivated by a 3D densely connected U-net, but requiring less memory usage, we developed a neural network architecture that combines the essence of these two influential neural network architectures into our proposed network while maintaining a respectable RAM usage, which we call Hierarchically Densely Connected U-net (HD U-net). The term "hierarchically" is used here to describe the different levels of resolution in the U-net between each max pooling or upsampling operation. The convolutional layers are densely connected along each hierarchy, but not between hierarchies of the U-net during the upsampling operation. In particular, we wish to utilize the global and local information capabilities of U-net and the more efficient feature propagation and reuse of DenseNet. DenseNet alone is not expected to perform well for this task as we conjecture that the accurate prediction of dose distribution requires both global and local information. While the feature maps of DenseNet are connected throughout the network, which allows for an efficient feature propagation, the lack of pooling followed by subsequent upsampling procedure, that is found in U-net, limits the network's capability to capture global information. In this study, we will assess the proposed deep learning architecture on its capability to volumetrically predict the dose distribution for patients with H&N cancer, and compare its performance against the two deep learning models from which it was inspired from: U-net and DenseNet. The HD U-net and the 3D variants of U-net and DenseNet can all fit on a 11GB 1080 Ti GPU for unbiased comparison.

## II. METHODS

### II.1. H&N Patient Data

We acquired a total of 120 H&N patients for this study. Table 1 summarizes some of the patient information. The H&N cancer sites included base of tongue, lateral border of tongue, tonsillar fossa, glottis, supraglottis, thyroid, thyroid gland, larynx, mouth, mandible, pharynx, oropharynx, nasopharynx, hypopharynx, pyriform sinus, tonsil, retromolar area, parotid gland, bone, laryngeal cartilage, sublingual gland, nasal cavity, vallecula, aryepiglottic fold, lingual tonsil, maxillary sinus, laryngeal cartilage, parts of face, and trachea. The specific ICD codes of the patients used this study included C01, C09.0, 161.0, 193, C79.89, C32.9, C06.9, C32.8, 161.9, C12, C73, C09.8, C06.2, C07, 145.6, 161.1, C11.3, C79.51, C77.0, C32.3, C08.1, D10.6, C02.3, C11.1, C09.9, C49.0, 160.0, C10.0, C13.1, 141.6, C31.0, 147.9, C10.8, 161.3, C11.2, 141.0, 149.0, 141.2, 160.2, 146.0, 172.3, 146.9, C33, C13.2, C02.1, C32.1, C76.0, C41.1, C02.8, and C06.89

|  | Min | Median | Mean | Max |
|---|---|---|---|---|
| Age (yr) | 18 | 60 | 61.43 | 91 |
| Number of targets | 1 | 3 | 2.94 | 5 |
| Prescription dose (Gy) | 42.5 | 60 | 61.51 | 72 |
| Total target volume (cc) | 38.75 | 700.06 | 706.73 | 1997.38 |
| Body volume (cc) | 10602.68 | 14451.15 | 13229.69 | 37060.00 |

Table 1: Summary of patient information

For each patient, we obtained the structure contours and the clinically delivered VMAT dose, calculated on the Eclipse Treatment Planning System. The voxel resolution of both the contours and dose were set to 5 mm$^3$. As input, each OAR was set as separate binary masks, where each voxel is assigned 1 if the voxel is assigned to the OAR and 0 otherwise, in their own channel. The patient CT was not included as input for this study. The PTVs were included as their own channel, but instead of a binary mask, the mask was set to have a value equal the prescribed radiation dose. Each patient had 1-5 PTVs, with prescription doses ranging from 42.5 Gy to 72 Gy. In total, the input data used 23 channels to represent the OARs, PTVs, and prescription doses. Because we wished to input the prescription dose for the PTV's in Gy and then output the resulting in dose distribution in Gy, we chose not to normalize the data. This means the prediction model will have to learn how to propagate the prescription information directly. The 22 OARs used in this study are the body, left and right brachial plexus, brain, brainstem, left and right cerebellum, left and right cochlea, constrictors, esophagus, larynx, mandible, left and right masseter, oral cavity, post arytenoid & cricoid space (PACS), left and right parotid, left and right submandibular gland (SMG), and spinal cord. In the case that the patient was missing one of the 22 OARs, the corresponding channel was set to 0 for the input.

## II.2. Deep Learning Terminology

In this section, we will define the some of the core deep learning terminology used in this paper. Convolutional neural networks (CNN) were first proposed in by LeCun et al [43,44], and had quickly found their use in deep learning for computer vision and imaging tasks. By using kernels and convolution, CNNs can easily extract image features, such as edges. Additionally, CNN feature extraction are shift invariant and use overall less weights than fully connected networks. Each convolution layer of a CNN calculates a set of feature maps from the input or a set of feature maps from the previous layer. DenseNet is a type of CNN, where the previous maps are concatenated to future layers.

Activation functions are non-linear functions typically used after a convolutional or fully connected operation. The Rectified Linear Unit (ReLU) [45] is one of today's most popular types of activation functions. It is defined as $f(x) = \max(0, x)$. It gained popularity for its resulting great performance in models. It did not suffer from the vanishing gradient issues found in other activation functions, such as the sigmoid or tanh functions. Also, it was computationally cheap to perform, which helped further speed up the training of neural networks.

Pooling in deep learning refers to dividing a feature map into rectangular patches, and then aggregating the pixel information in each patch to create a new, lower-resolution layer that retains important features of the high resolution map. Typically, the patches are 2 x 2 for 2D and 2 x 2 x 2 for 3D inputs with a stride of 2, which effectively halves each dimension on the output feature map. This greatly reduces the computational expense and helps the network see the image more globally while using a standard size convolution kernel. Max pooling [46], where from each rectangular patch, the single largest pixel value is carried over, became one of the most popular pooling methods to use with CNNs. There is no direct inverse operation to max pooling, but some common techniques to increase the resolution include upsampling and deconvolution [47].

The loss or cost function is a term used in both machine learning and optimization, and is a function that quantifies a particular state into a single value. The goal is to minimize or maximize this value by changing a set of defined parameters in the problem. For deep learning, the network is parameterized by a large number of weights, and the loss, for supervised learning, is typically defined as some type of error between the neural network's prediction and a known ground truth. By being fed many examples, the network's weights are tuned to minimize this error. Some very common loss functions used in deep learning today are mean squared error and cross entropy.

## II.3. Deep Learning Architectures

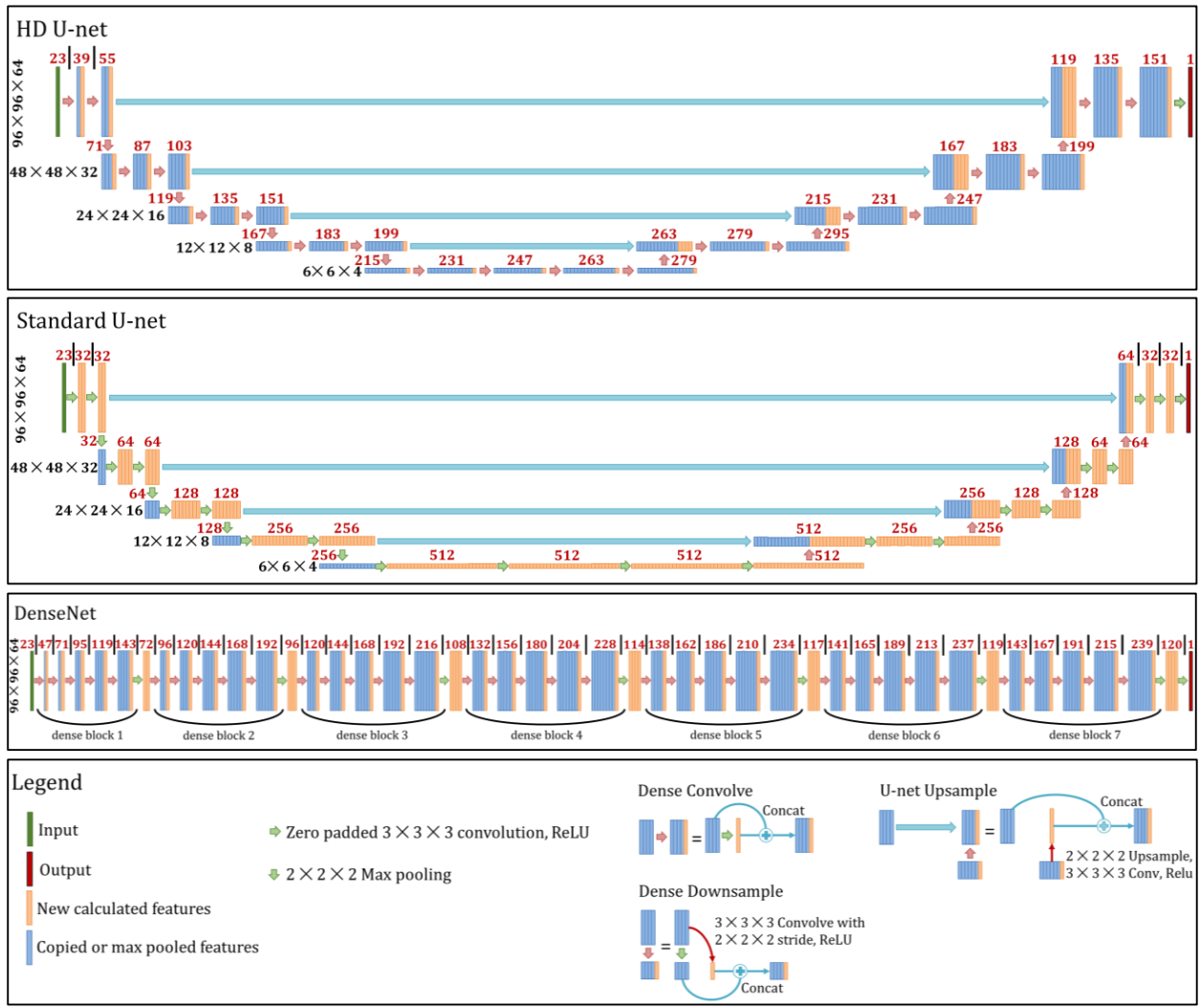

Figure 1: Architectures used within the study. Black numbers on the left side of the model represent the volume shape and resolution at a specific hierarchy. Red numbers represent the number of feature maps at a particular layer. Orange features represent the newly calculated features, and trainable parameters to learn, while blue features are copied or max pooled features that do not need trainable parameters.

Figure 1 shows all of the specific architectures that were used in the study. The HD U-net utilizes 3 operations defined in the legend: dense convolve, dense downsample and U-net upsample. The dense convolve effectively uses a standard convolution with ReLU, followed by a concatenation of the previous feature set. Performing this operation back-to-back is equivalent to the densely connected computation in the DenseNet publication. The dense downsampling operation is performed by a strided convolution and ReLU to calculate a new feature set that has half the resolution. The previous feature set is then max pooled and concatenated to the new feature set. Lastly, the U-net upsampling operation is done by up-sampling, convolution, and ReLU, followed by a concatenation of the feature set on the other side of the "U". This "u-net upsampling" is the same operation used in the standard U-net, with the upsample + convolve sometimes replaced with the transposed convolution. For each dense operation, a growth rate can be defined as the number new features calculated during the convolution step. Specifically, we utilized a growth rate of 16 (16 new features added after each "dense" operation), 4 dense downsampling operations, and 64 features returned during the upsampling operation.

To assess the performance of our implementation, our Hierarchically Dense (HD) U-net is compared to the two models which had inspired its design: the standard U-net and DenseNet. To fairly assess the architectures, the standard U-net, was constructed to match the HD U-net in terms of the number of downsampling operations used, and followed the conventional build of U-net, where the number of filters are doubled after each max pooling operation. It utilizes the regular convolution and max pooling, as defined in the green arrows in the legend in Figure 1.

DenseNet was constructed as outlined in the DenseNet publication, with dense-blocks followed by compression layers. Since DenseNet has an entirely different architecture than the U-nets, we simply chose to match the number of trainable parameters to HD U-net as close as possible. Ultimately, to meet this number of parameters, we chose to have 7 dense blocks, 5 dense convolutions per dense block, and a compression factor of 0.5. DenseNet utilizes the dense convolution during the dense block, and a normal convolution that reduces the number of layers by our set compression factor.

All networks were constructed to use 3D operations to handle the volumetric H&N data. After determining that overfitting was not occurring, based on the training the validation loss curves shown in Figure 2, it was decided that dropout, L1 regularization, and L2 regularization will not be used in the final model. Batch normalization was briefly tested, but after some trial-and-error testing, it was ultimately removed in the final models. Exact details of each network is summarized in the appendix.

### II.4. Training and Evaluation

Of the 120 H&N patients, we set aside 20 patients as testing data to evaluate at the end. To assess the performance and stability of each model—HD U-net, Standard U-net, and DenseNet—a 5-fold cross validation procedure was performed on the remaining 100 patients, where, for each fold, the patients were divided into 80 training patients and 20 validation patients. During each fold, the model would have its weights randomly initialized, and then update its weights based on the training set. The validation loss is used to determine the iteration that had the best model weights, a well-known method to help prevent from obtaining a model that has overfitted to the data[48,49]. This instance of the model is then used to evaluate the validation data. After all models from every fold was trained, the models then evaluated the testing data, and the results combined and averaged through bootstrap aggregation (bagging) [50] without replacement [51]. Bagging an ensemble of predictors helps improve the stability and accuracy of the algorithm, and aids in preventing overfitting.

Mean squared error between the predicted dose and the clinically delivered dose was used as the loss function for training each neural network model. The learning rate of each model was adjusted to minimize the validation loss as a function of epochs. The patch size used for neural training was 96 x 96 x 64. Matching the U-net paper[37], we used a batch size of 1 sample. Instead of subdividing the patient contours and their corresponding dose volumes into set patches, each iteration of the model training process randomly selected a patch from the patient volume on-the-fly. Random patch selection also inherently adds stochastic translational shifts into the training data, which is one of the basic forms of data augmentation. While not the most robust data augmentation, this helps to somewhat reduce the overfitting issue, particularly for small datasets.

To equally compare across the patients, all plans were normalized such that the PTV with the highest corresponding prescription dose had 95% of its volume receiving 100% of the prescription dose ($D95$). All dose statistics will also be reported relative to the prescription dose (i.e. the errors are reported as a percent of the prescription dose). As evaluation criteria PTV coverage (D98, D99), PTV max dose, homogeneity $\left(\frac{D2-D98}{D50}\right)$, van't Riet conformation number[52] $\left(\frac{(V_{PTV} \cap V_{100\%Iso})^2}{V_{PTV} \times V_{100\%Iso}}\right)$, and the structure max and mean doses ($D_{max}$ and $D_{mean}$) were evaluated.

To maintain consistency in performance, all neural network models were trained and evaluated on an NVIDIA GTX 1080 Ti GPU with 11 GB dedicated RAM.

## III. RESULTS

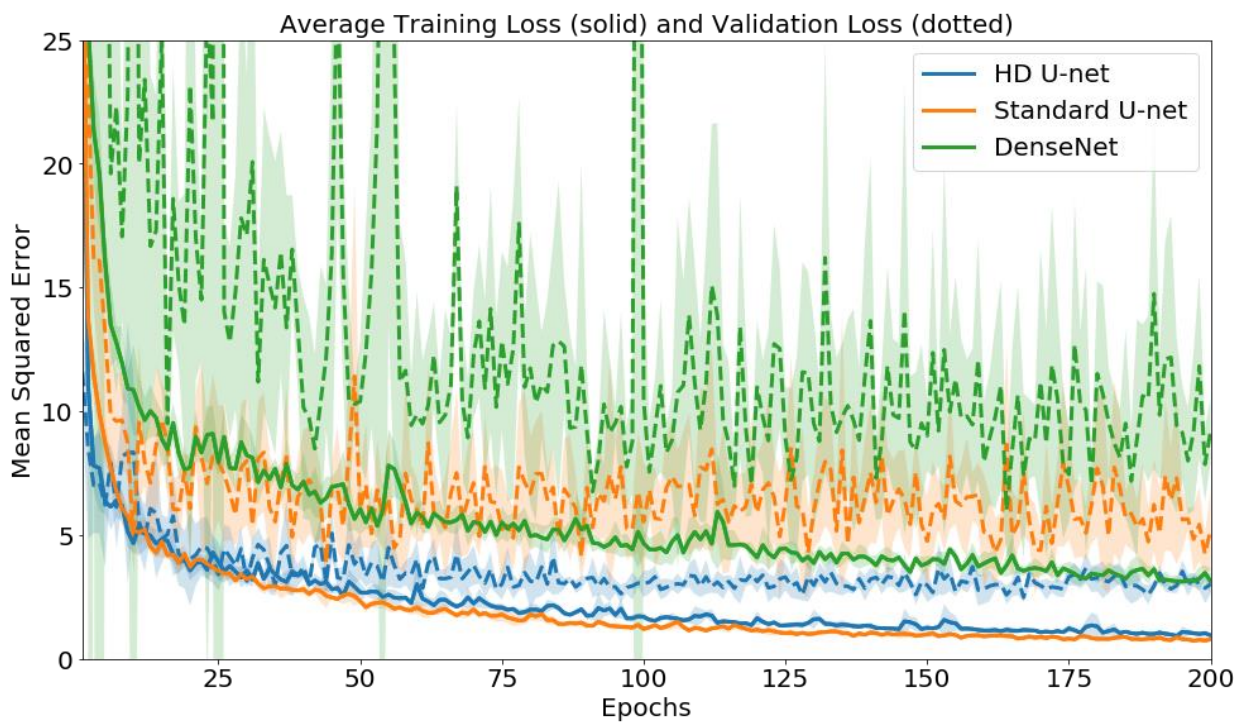

Figure 2: Mean loss across the 5 cross-validation folds for each model. The error equals 1 standard deviation.

Figure 2 shows the mean training and validation loss for the HD U-net, Standard U-net, and DenseNet. The HD and Standard U-net have a similar training loss as a function of epochs. However, the validation loss of the HD U-net is much lower and has less variation between the folds of the cross validation than that of the Standard U-net. This indicates that HD U-net is better at generalizing the modeling contours-to-dose, and is overfitting less to the training data. The DenseNet performed the worst for both the mean training and validation loss, as well as having the largest variation in the validation loss.

|  | Trainable parameters | Training Time (s) (averaged across cross validation folds) Mean ± SD | Prediction time for entire patient volume (s) Mean ± SD | |
|---|---|---|---|---|
|  |  |  | Test | Cross-Val |
| HD U-net | 3,289,006 | 11138.4 ± 50.6 | 5.42 ± 1.99 | 5.39 ± 2.39 |
| Standard U-net | 40,068,385 | 8688.0 ± 50.2 | 4.48 ± 1.67 | 4.60 ± 2.01 |
| DenseNet | 3,361,708 | 31764.0 ± 53.7 | 17.12 ± 6.42 | 18.05 ± 7.97 |

Table 2: Trainable parameters and prediction time for each model.

Table 2 shows the number of trainable parameters and the prediction time for each model used in the study. The HD U-net and DenseNet have approximately 12 times less trainable parameters than the Standard U-net. The training time of HD U-net was about 3 hours, which is slightly longer than the Standard U-net's time of 2.4 hours. DenseNet had the longest training time of 8.8 hours. The prediction time of the HD U-net is approximately 1 second longer for a full patient prediction, using patches of 96 x 96 x 64 and stride of 48 x 48 x 32. DenseNet had the longest prediction time of about 4 times longer than either of the U-nets.

| | | PTV dose coverage and max dose | | | |
|---|---|---|---|---|---|
| | | Average values $\left(\frac{1}{n}\sum_{i=1}^{n}\frac{Value_i}{Prescription\ Dose_i}\right)$ | | | |
| | | D95 | D98 | D99 | $D_{max}$ |
| | | Mean ± SD | Mean ± SD | Mean ± SD | Mean ± SD |
| Test Results | Ground Truth | 0.9999 ± 0.0002 | 0.99 ± 0.02 | 0.97 ± 0.03 | 1.04 ± 0.02 |
| | HD U-net | 0.9999 ± 0.0005 | 0.99 ± 0.01 | 0.97 ± 0.02 | 1.07 ± 0.02 |
| | Standard U-net | 1.0000 ± 0.0006 | 0.98 ± 0.02 | 0.96 ± 0.03 | 1.11 ± 0.03 |
| | DenseNet | 0.9900 ± 0.0993 | 0.97 ± 0.10 | 0.96 ± 0.11 | 1.11 ± 0.16 |
| Cross-Validation Results | Ground Truth | 1.0002 ± 0.0003 | 0.96 ± 0.07 | 0.91 ± 0.17 | 1.06 ± 0.04 |
| | HD U-net | 1.0001 ± 0.0004 | 0.98 ± 0.03 | 0.94 ± 0.11 | 1.08 ± 0.03 |
| | Standard U-net | 0.9999 ± 0.0005 | 0.98 ± 0.01 | 0.95 ± 0.14 | 1.13 ± 0.04 |
| | DenseNet | 1.0001 ± 0.0004 | 0.98 ± 0.02 | 0.95 ± 0.11 | 1.09 ± 0.04 |

Table 3: Average PTV coverage and max dose for the ground truth and the prediction models.

| | | PTV dose coverage and max dose | | | | | | | |
|---|---|---|---|---|---|---|---|---|---|
| | | Average percent prediction error $\left(\frac{1}{n}\sum_{i=1}^{n}\frac{|Truth_i-Prediction_i|}{Prescription\ Dose_i}\times 100\right)$ | | | | | | | |
| | | D95 | | D98 | | D99 | | $D_{max}$ | |
| | | Mean ± SD | p-val | Mean ± SD | p-val | Mean ± SD | p-val | Mean ± SD | p-val |
| Test Results | HD U-net | 0.02 ± 0.05 | — | 1.18 ± 1.82 | — | 1.96 ± 2.14 | — | 3.75 ± 1.60 | — |
| | Standard U-net | 0.03 ± 0.06 | 0.261 | 1.77 ± 2.35 | 0.001 | 2.65 ± 2.95 | 0.006 | 7.42 ± 3.26 | 2.54 x $10^{-17}$ |
| | DenseNet | 1.01 ± 9.93 | 0.322 | 2.45 ± 10.13 | 0.207 | 3.42 ± 10.39 | 0.148 | 7.26 ± 15.37 | 0.026 |
| | HD U-net | 0.02 ± 0.05 | — | 2.69 ± 6.13 | — | 6.02 ± 12.94 | — | 3.84 ± 3.13 | — |
| Cross-Validation Results | Standard U-net | 0.03 ± 0.06 | 0.365 | 2.86 ± 6.50 | 0.423 | 5.50 ± 9.69 | 0.651 | 7.64 ± 4.33 | 1.01 x 10⁻¹² |
| | DenseNet | 0.03 ± 0.06 | 0.087 | 2.82 ± 6.32 | 0.473 | 6.41 ± 13.59 | 0.259 | 5.02 ± 3.98 | 1.20 x 10⁻⁴ |

Table 4: PTV coverage and max dose prediction errors for each model. The p-values < 0.05 signify the statistical significance of the difference in the errors of the predictions of HD U-net against the Standard U-net and DenseNet.

| Homogeneity and van't Riet conformation number | | | |
|---|---|---|---|
| Average values $\left(\frac{1}{n}\sum_{i=1}^{n} Value_i\right)$ | | | |
| | | Homogeneity $\left(\frac{D2-D98}{D50}\right)$ | van't Riet conformation number |
| | | Mean ± SD | Mean ± SD |
| Test Results | Ground Truth | 0.06 ± 0.04 | 0.78 ± 0.06 |
| | HD U-net | 0.08 ± 0.02 | 0.76 ± 0.06 |
| | Standard U-net | 0.13 ± 0.04 | 0.74 ± 0.07 |
| | DenseNet | 0.12 ± 0.18 | 0.74 ± 0.12 |
| Cross-Validation Results | Ground Truth | 0.09 ± 0.09 | 0.74 ± 0.09 |
| | HD U-net | 0.10 ± 0.05 | 0.73 ± 0.08 |
| | Standard U-net | 0.14 ± 0.04 | 0.73 ± 0.08 |
| | DenseNet | 0.10 ± 0.04 | 0.74 ± 0.07 |

Table 5: Average homogeneity and van't Riet conformation numbers for the ground truth and the prediction models.

| Homogeneity and van't Riet conformation number | | | | | |
|---|---|---|---|---|---|
| Average percent prediction error $\left(\frac{1}{n}\sum_{i=1}^{n} \frac{|Truth_i - Prediction_i|}{1} \times 100\right)$ | | | | | |
| | | Homogeneity $\left(\frac{D2-D98}{D50}\right)$ | | van't Riet conformation number | |
| | | Mean ± SD | p-val | Mean ± SD | p-val |
| Test Results | HD U-net | 3.74 ± 1.78 | — | 3.08 ± 2.30 | — |
| | Standard U-net | 7.39 ± 3.99 | 1.20 x 10⁻³ | 4.73 ± 4.30 | 3.37 x 10⁻⁹ |
| | DenseNet | 7.05 ± 17.39 | 0.063 | 6.28 ± 10.56 | 0.029 |
| | HD U-net | 5.25 ± 6.76 | — | 4.84 ± 4.08 | — |
| Cross-Validation Results | Standard U-net | 8.10 ± 6.04 | 3.09 x 10$^{-7}$ | 6.12 ± 5.73 | 0.307 |
| | DenseNet | 6.11 ± 6.90 | 0.004 | 6.70 ± 5.21 | 0.357 |

Table 6: Homogeneity and van't Riet conformation number prediction errors, taken as a percentage of the maximum of its achievable range. In our case, both homogeneity and the conformation number are defined to range from 0 to 1. The p-values < 0.05 signify the statistical significance of the difference in the errors of the predictions of HD U-net against the Standard U-net and DenseNet.

Table 3 shows the average values of the PTV coverage and max dose of the ground truth dose and the predicted clinical dose, and Table 4 shows the percent errors in the models' prediction on PTV coverage and max dose. While the models had similar performance in D95, D98 and D99 for the cross-validation data, the HD U-net had statistically significantly better performance in predicting the dose coverage on the test set, as compared to that of Standard U-net. HD U-net also had significantly less error in predicting the maximum dose to the PTV, as compared to the other two networks. Table 5 reports the average homogeneity indices and the van't Riet conformation numbers for the clinical dose and the predicted dose from the networks, and Table 6 shows the percent errors in the model's prediction of homogeneity and van't Riet conformation numbers. Statistically, the HD U-net's prediction on the test set of data has significantly less error in homogeneity and conformation than that of the Standard U-net, as well as significantly less error in conformation that that of the DenseNet. In addition, for the cross validation, HD U-net prediction had significantly less error than the other two network's predictions in homogeneity.

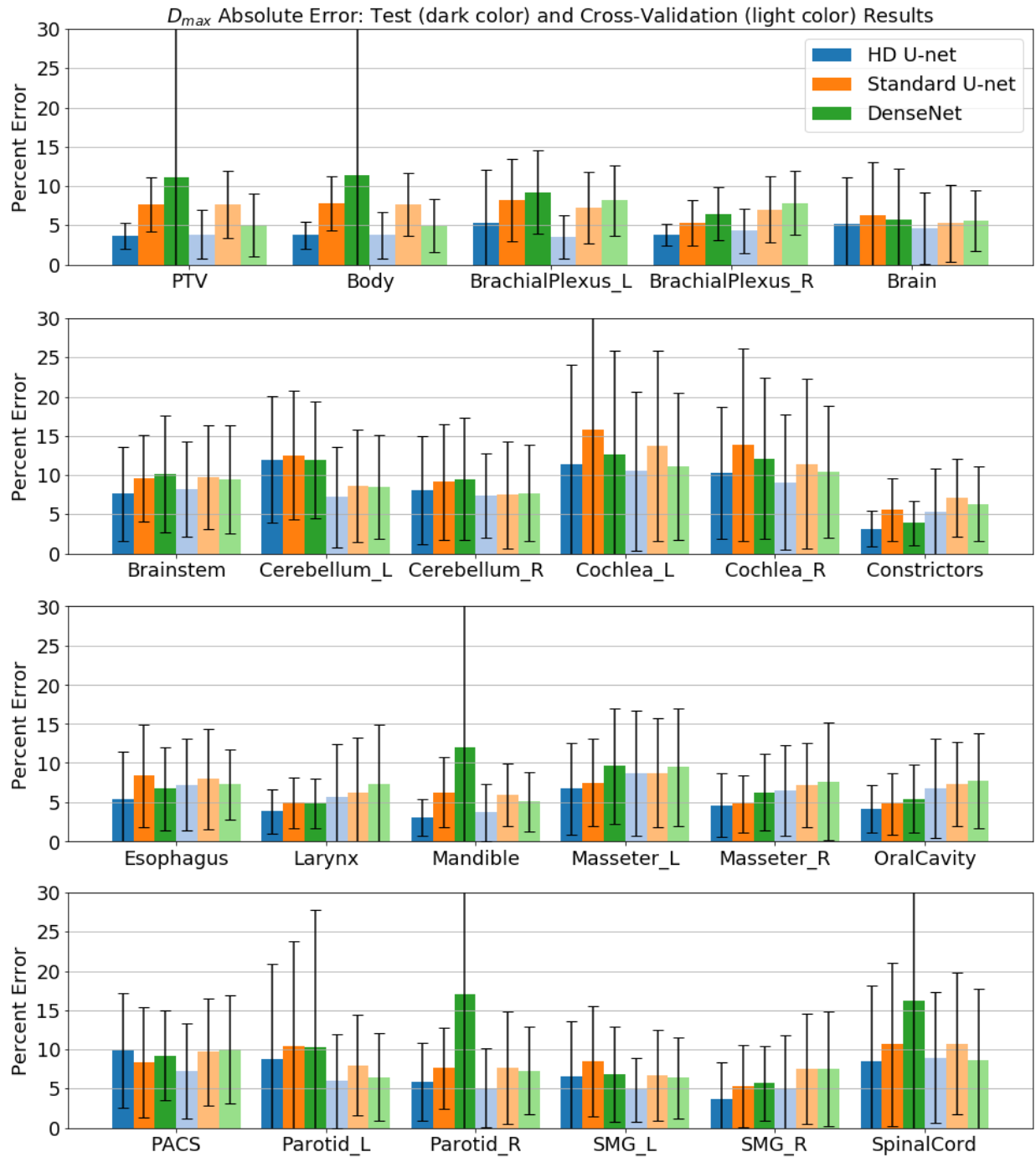

Figure 3: Absolute Error of $D_{max}$ on the structures of interest. Error is reported as a percentage of the prescription dose.

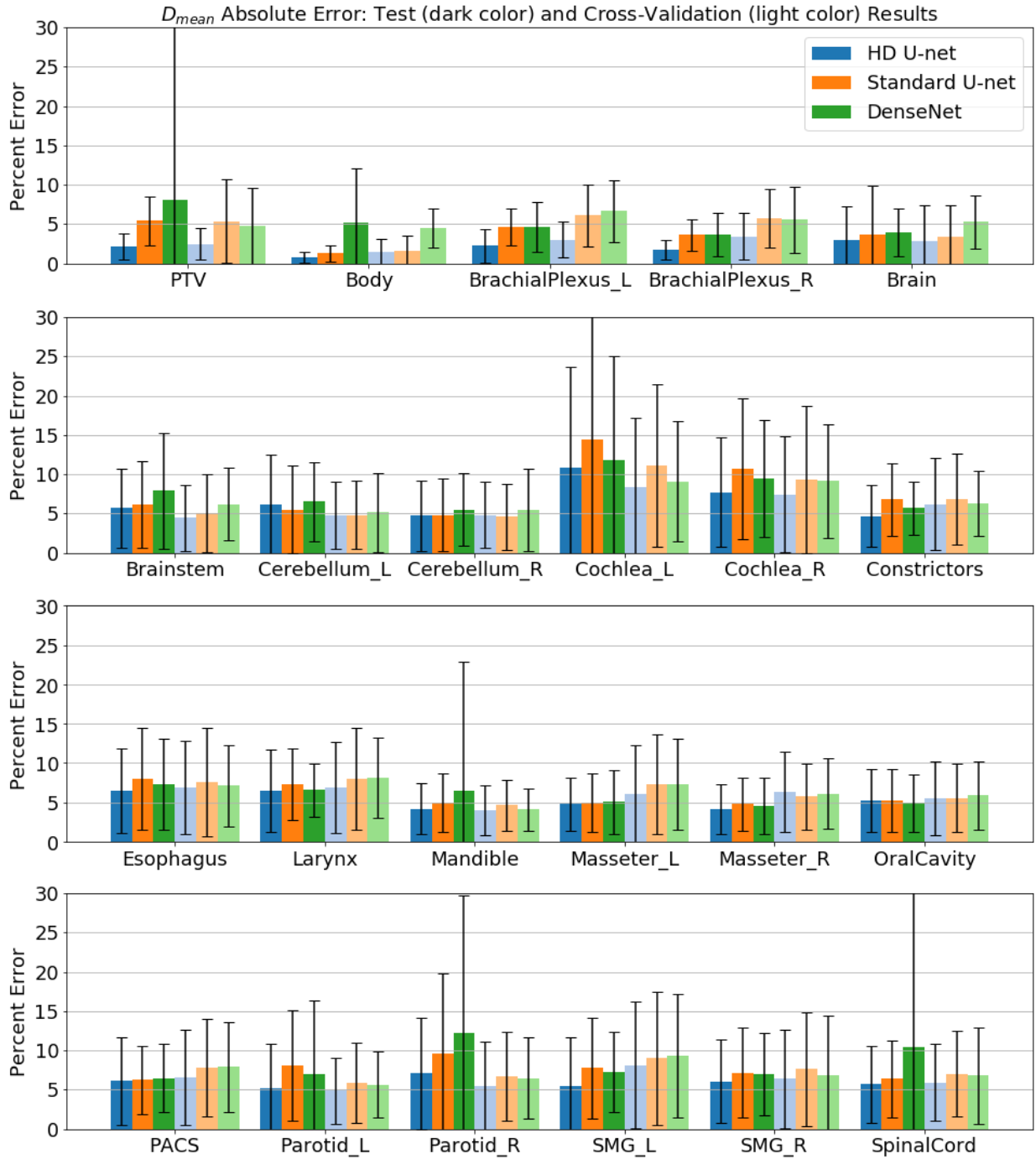

Figure 4: Absolute Error of $D_{mean}$ on the structures of interest. Error is reported as a percentage of the prescription dose.

Figure 3 and Figure 4 show the $D_{max}$ and $D_{mean}$ absolute errors on all of the 22 structures and PTV. Due to the large variability in number of PTVs and prescription doses, percent errors are reported as a percent of the highest prescription dose for the patient, and the PTV $D_{mean}$ and $D_{max}$ calculation for Figure 3 and Figure 4 used the union of all the plan's PTVs as the region of interest. Normalization by the prescription dose allows for one to know how much error to expect given a prescription, and can decide if more attention to an OAR is needed. It can be easily seen that the HD U-net, shown in blue, has an overall lower prediction error on the $D_{max}$ and $D_{mean}$ than the other two networks in this study. For the cross-validation data, the HD U-net, Standard U-net, and DenseNet predicted, on average, the $D_{max}$ within 6.23 ± 1.94%, 8.11 ± 1.87%, and 7.65 ± 1.67%, respectively, and the $D_{mean}$ within 5.30 ± 1.79%, 6.38 ± 2.01%, and 6.49 ± 1.43%, respectively, of the prescription dose. For the test data, the models predicted $D_{max}$ within 6.30 ± 2.70%, 8.21 ± 2.87%, and 9.30 ± 3.44%, respectively, and $D_{mean}$ within 5.05 ± 2.13%, 6.40 ± 2.63%, and 6.83 ± 2.27%, respectively, or the prescription dose. Overall, the HD U-net had the best performance on both the cross-validation and test data. DenseNet had the largest discrepancy between its performance on the cross-validation data and test data, indicating its prediction volatility on data outside of its training and validation set.

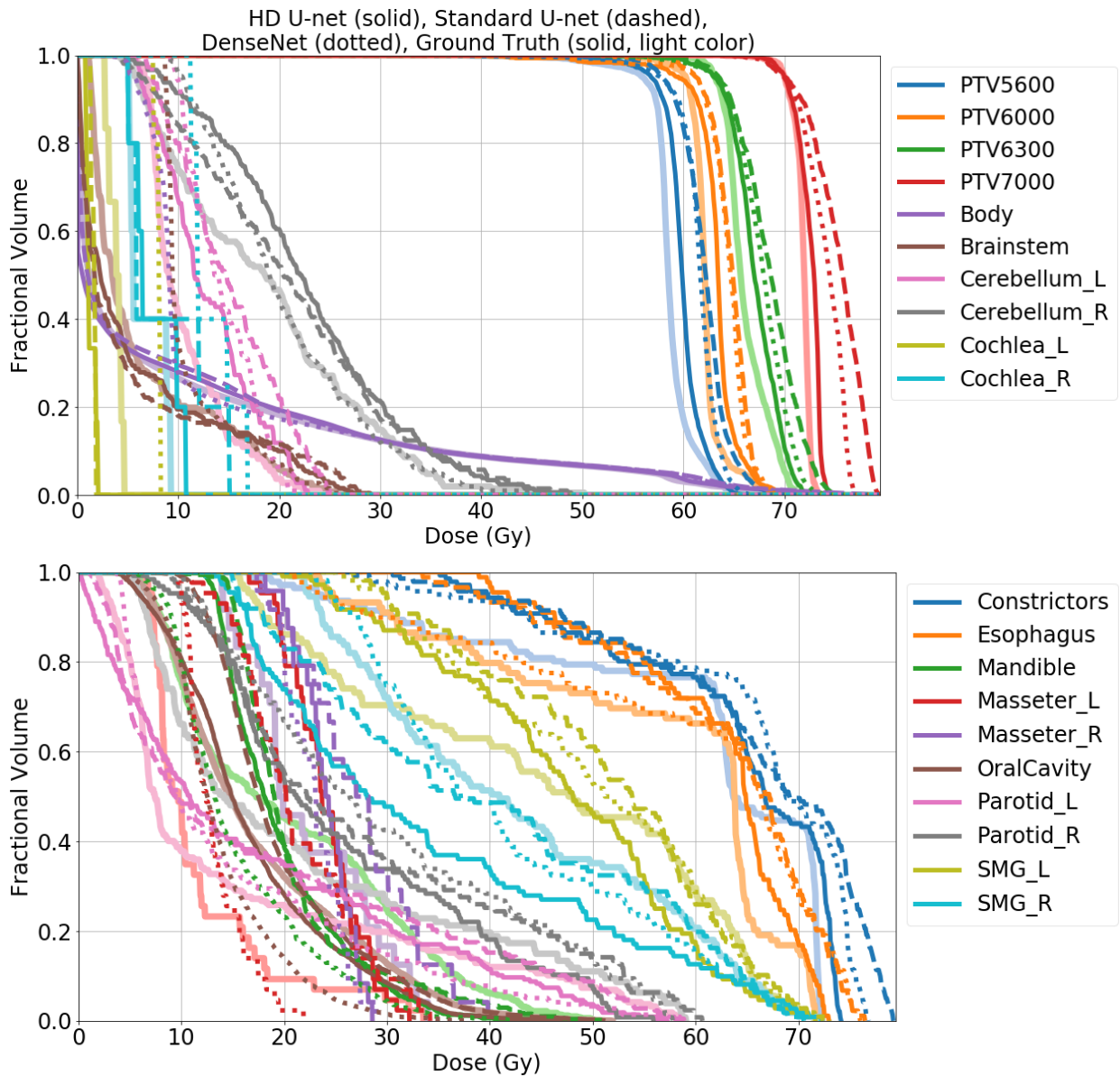

Figure 5: Dose volume histogram of an example patient from the test pool.

Figure 5 shows an example DVH from a patients from the test data. The solid line with the lighter color variant represents the clinical ground truth dose, while the darker color variants represent the predicted dose from HD U-net (solid), Standard U-net (dashed), and DenseNet (dotted). For this example patient, the HD U-net is superior to the other models in predicting the dose to the PTVs. Prediction of OAR dose are more variable between the models. This is also reflected in Figure 3 and Figure 4, where the standard deviation in prediction is small for the PTVs using the HD U-net, and larger on the OAR $D_{max}$ and $D_{mean}$ prediction.

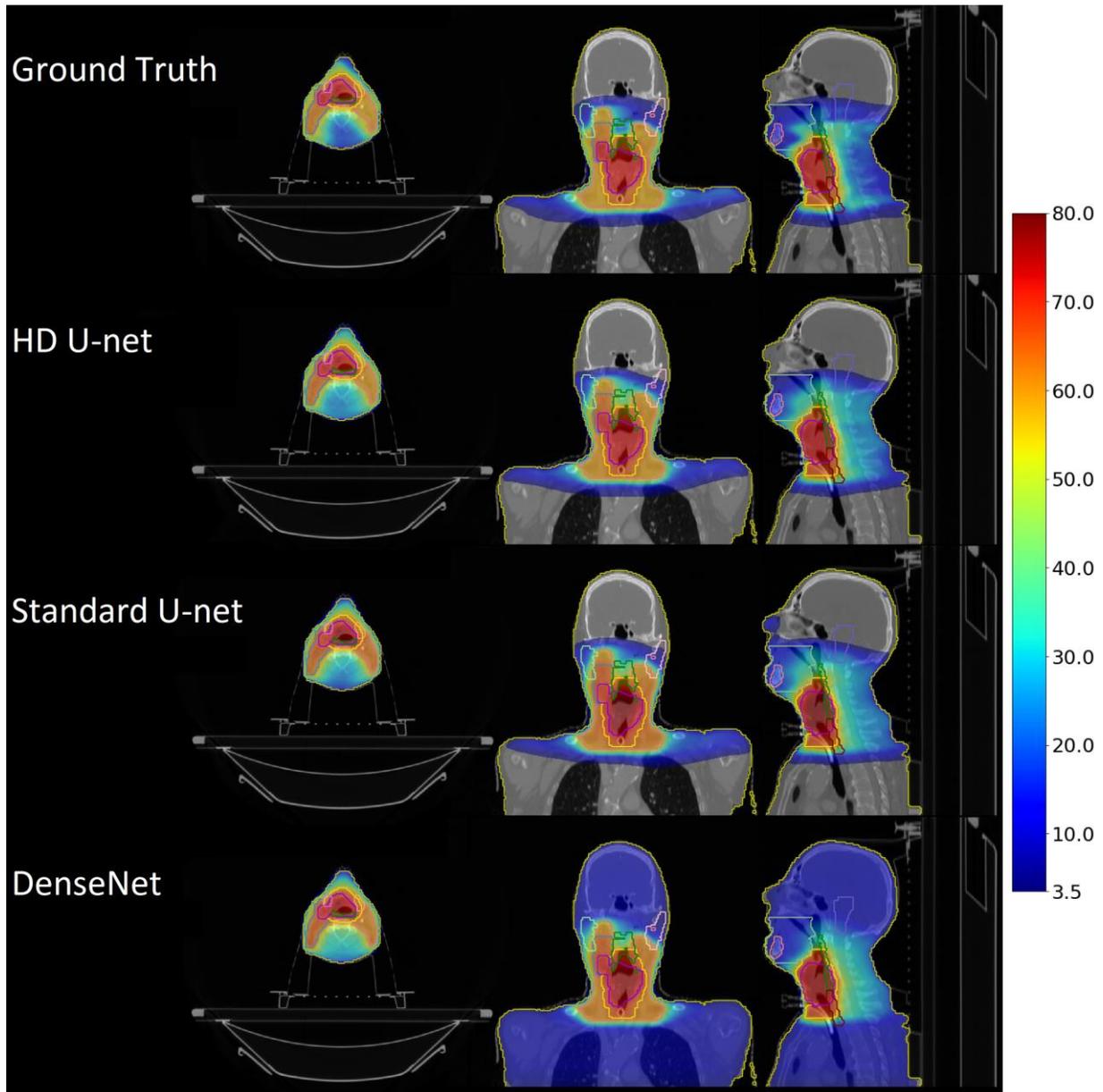

Figure 6: Dose washes of example patient from the test pool. The color bar is shown in units of Gy. The clinical ground truth dose is shown on the top row, followed by the dose

predictions of the HD U-net, Standard U-net, and DenseNet, respectively. Low dose cutoff for viewing was chosen to be 5% of the highest prescription dose (3.5 Gy).

Figure 6 shows the dose color wash for the same patient in Figure 5. Visually, the dose prediction models have comparable dose to the PTVs, with the Standard U-net and DenseNet slightly hotter than the HD U-net. The DenseNet also predicts dose above 3.5 Gy everywhere in the body, which is also reflected in the DVH in Figure 5 (purple dotted line) and the dose wash in Figure 6. The back of the neck is predicted to have more dose by all of the models, as compared to ground truth, which may represent a lack of data representation in the training data, or a lack of information being fed into the deep learning model itself.

## IV. DISCUSSION

To our knowledge, this is the first instance of an accurate volumetric dose prediction for H&N cancer patients treated with VMAT. Existing plan prediction models are largely based around Knowledge Based Planning (KBP)[53-66], with clinical/commercial implementations available known as Varian RapidPlan (Varian Medical Systems, Palo Alto, CA) and Pinnacle Auto-Planning Software (Philips Radiation Oncology Systems). These KBP methods have historically been designed to predict the DVH of a given patient, instead of the full volumetric dose prediction. The only exception is the study by Shiraishi and Moore[58] in 2016, where they perform 3D dose prediction. However, their study is currently only evaluated on prostate patients, and thus the results are not comparable to our results for H&N patients. A study by Tol et al.[67] that evaluated RapidPlan on H&N cancer patients, had found that, in one of their evaluation groups, RapidPlan, had a mean prediction error of as large as 5.5 Gy on the submandibular gland, with the highest error on a single patient's OAR as high as 21.7 Gy on the lower larynx. Since their patients were clinically treated from 54.25 to 58.15 Gy, this translates to roughly 10% and 40% error, respectively in predictive performance. Another study by Krayenbuehl et al.[68] had used Pinnacle Auto-Planning Software. However, in this study, the plan prediction aspect of the software was hidden from the user, and simply used as part of the auto-planning software itself, making this study's methodology not directly comparable to ours.

As a future study, we intend to perform a comprehensive evaluation and comparison of our dose prediction model against Varian RapidPlan. RapidPlan analyzes distance-to-target histograms—a relative geometrical relationship between an OAR and the PTV. In addition, RapidPlan incorporates relative overlap volume, relative out-of-field volume, absolute OAR volume, and absolute target volume[69]. Unlike our deep learning model, RapidPlan does not consider relationships between the OARs themselves, and thus does not account for tradeoffs between them. For an impartial comparison, both the RapidPlan and our dose prediction model must be trained and tested on the same data set. We plan to acquire a clean dataset that was used to train RapidPlan to accomplish an unbiased study.

It is currently a challenge to directly compare against other non-commercial prediction models, particularly since they are developed in-house and are proprietary to the institution that developed it. It is typically infeasible to obtain a copy or to faithfully replicate it to the exact specifications that were used by the originators. In addition, training and evaluation of the model is usually performed using the institution's own data, and is often unavailable to the public to replicate the results or to train their own model for an unbiased comparison. These are common issues in the era of data driven modeling for the field of medicine, and while efforts to develop large public datasets exist, we have yet to reach the point where it is commonplace to use communal datasets for apples-to-apples comparisons.

Although the DenseNet had the poorest performance of the 3 models, it is due to the fact that the DenseNet is incapable of capturing global information into its prediction as the U-nets are capable of. This should not be seen as an oversight of DenseNet, as the authors of the paper proposed the concept of densely connected convolutional neural networks as a module, implying that this concept can be applied to more complex models. Their proposed DenseNet was used to illustrate the efficient feature propagation and reuse, alleviate the vanishing gradient, and reduce the number of parameters to moderate the overfitting issue. While DenseNet also had the slowest training and prediction time, it is possible that this issue may be alleviated with the newer GPUs, that are designed specifically for deep learning, such as the Volta architecture with tensor cores.

While CT information was not used as an input, the ground truth clinical plan doses were calculated using the Eclipse dose calculation engine, which does account for CT information for the inhomogeneity correction. Since we are training the neural network to learn to predict this particular dose from just contours, the deep neural network may learn the correction itself without the need of the CT information. We plan to fully evaluate the effects of adding the CT as an additional input into the model in a future study.

While accuracy tends to be positively correlated to a specific number of parameters in neural networks, the relationship is still not entirely clear between the accuracy and number of parameters, since neural networks are highly non-linear functions. Prediction time also tends to increase with more parameters and higher memory utilization, but is also affected by the breadth and depth of the network. Given the same number of trainable parameters, a fatter network with less layers will tend to predict more quickly than a skinny network with many layers, since the GPU can take advantage of the parallel computation.

The current main disadvantage of the densely connected architecture is that it requires a much larger memory usage, per trainable parameter, than standard networks. This is because it carries many of the past features it has calculated in memory in order to compute the next feature map. Xiaomeng et al. previously attempted to combat this with a hybrid densely connected U-net, where they combine a 2D Dense U-net, to extract intra-slice features, with a 3D counterpart for aggregating volumetric contexts[70]. However, on a 12 GB RAM Titan Xp, the number of input slices they used at a time is 12, with feature maps containing only 3 slices during the dense block calculation. For radiotherapy, we require a much larger number of slices (e.g. 64) to make an accurate dose prediction, particularly for a complex site such as H&N. This makes the hybrid dense U-net by Xiaomeng et al. infeasible to use since it no longer fits on a modern GPU, hence our motivation for a less RAM intensive version of a densely connected U-net. Even with the extra memory cost, the expense is offset by the exceptionally efficient feature calculation. The Standard U-net, while capable of utilizing both the local and global information into its prediction still has considerable prediction error, indicating that either the model is not intelligent enough for the task, or it is not using its trainable parameters efficiently enough. In this study, we have shown, that by constructing a U-net with densely connected properties, we can take advantage of both U-net and DenseNet in our proposed model, and we are capable of reducing the total number of trainable parameters by 12 fold while attaining a superior prediction result, as compared to the Standard U-net.

There are currently several limitations to the study and the resulting model. One apparent limitation is the data size. Typical deep learning algorithms use considerably larger datasets in the hundreds of thousands or millions if needed. While we did not observe overfitting issues in the training and validation curves, possibly due to the fully convolutional nature of the U-net architecture and the relatively tight scope of the project goal, having substantially larger dataset may further tighten the generalization gap (the gap between the training and validation loss), and improve predictive performance. Another limitation is the inflexibility of the model with respect to adding more OARs. We trained the model to have the PTVs and 22 specific structures as input. While the model can handle dose prediction on patients that are lacking some of the 22 structures, there is not a way for the model to account for other structures without entirely retraining the model.

This dose prediction tool can currently be used as a clinical guidance tool, where the final tradeoff decisions and deliverable plan will still be made by the physician and dosimetrist. The amount of prediction error of the model can be used alongside the prediction during clinical guidance. As an average between all OARs, the HD U-net model can predict the OAR Dmax and Dmean within 6.3% and 5.1% of the prescription dose. It is quite possible that the model predicts for the dose of a particular OAR to be within absolute constraints, but is unachievable when the treatment plan is created. If the difference between the predicted values and the constraint is within the prediction error, the OARs of interested can be highlighted to be focused on during the treatment planning phase.

We plan to expand on this study and improve the model in several ways. First, to further reduce the prediction error to the OARs, we plan to incorporate the dose constraints, as prescribed by the physician, an input into the prediction model. Furthermore, we will examine the addition and effects of the CT image as input on the prediction accuracy. We expect that the addition of these types of information will prominently improve the deep learning model's performance, and will investigate to quantify their impact. In addition, we plan to develop and integrate a dose mimicking optimization [71] to convert our predicted dose into the best machine parameters to deliver the plan. As the field of artificial intelligence and deep learning continues to explosively progress, we will persistently investigate new deep learning concepts and architectures for more intelligent and efficient models to use. Eventually, our goal is to transition away from dose prediction based on historical plans to an AI-based treatment planning system, where we can push radiotherapy plans to improve and become better than that from current or past clinical practices.

## V. CONCLUSION

We have developed and proposed a hierarchically densely connected U-net architecture, HD U-net, and applied the model to volumetric dose prediction for patients with H&N cancer. Using our proposed implementation, we are capable of accurately predicting the dose distribution from the PTV and OAR contours, and the prescription dose. On average, our proposed model is capable of predicting the OAR max dose within 6.3% and mean dose within 5.1% of the prescription dose on the test data. The other models, the Standard U-net and DenseNet, performed worse, having an OAR max dose prediction error of 8.2% and 9.3%, respectively, and mean dose prediction error of 6.4% and 6.8%, respectively. HD U-net also outperformed the other two models in homogeneity, dose conformity, and dose coverage on the test data. In addition, the proposed model is capable of using 12 times less trainable parameters than the Standard U-net, and predicted the patient dose 4 times faster than DenseNet. We plan to continue improving the model by incorporating more dose constraints and evaluating the addition of other data, such as the patient CT, on the model's predictive performance. Our long-term goal is the development of an artificially intelligence-based radiotherapy planning system.

## VI. ACKNOWLEDGEMENTS

We would like to thank the support from Cancer Prevention and Research Institute of Texas (CPRIT) (IIRA RP150485, MIRA RP160661).

# VII. REFERENCES

1. Brahme, A. Optimization of stationary and moving beam radiation therapy techniques. *Radiotherapy and Oncology* **12**, 129-140 (1988).
2. Bortfeld, T., Bürkelbach, J., Boesecke, R. & Schlegel, W. Methods of image reconstruction from projections applied to conformation radiotherapy. *Physics in Medicine and Biology* **35**, 1423 (1990).
3. Bortfeld, T. R., Kahler, D. L., Waldron, T. J. & Boyer, A. L. X-ray field compensation with multileaf collimators. *International Journal of Radiation Oncology* Biology* Physics* **28**, 723-730 (1994).
4. Webb, S. Optimisation of conformal radiotherapy dose distribution by simulated annealing. *Physics in Medicine and Biology* **34**, 1349 (1989).
5. Convery, D. & Rosenbloom, M. The generation of intensity-modulated fields for conformal radiotherapy by dynamic collimation. *Physics in Medicine and Biology* **37**, 1359 (1992).
6. Xia, P. & Verhey, L. J. Multileaf collimator leaf sequencing algorithm for intensity modulated beams with multiple static segments. *Medical Physics* **25**, 1424-1434, doi:doi:http://dx.doi.org/10.1118/1.598315 (1998).
7. Keller-Reichenbecher, M.-A. *et al.* Intensity modulation with the "step and shoot" technique using a commercial MLC: A planning study. *International Journal of Radiation Oncology* Biology* Physics* **45**, 1315-1324 (1999).
8. Yu, C. X. Intensity-modulated arc therapy with dynamic multileaf collimation: an alternative to tomotherapy. *Physics in Medicine and Biology* **40**, 1435 (1995).
9. Otto, K. Volumetric modulated arc therapy: IMRT in a single gantry arc. *Medical physics* **35**, 310-317 (2008).
10. Palma, D. *et al.* Volumetric Modulated Arc Therapy for Delivery of Prostate Radiotherapy: Comparison With Intensity-Modulated Radiotherapy and Three-Dimensional Conformal Radiotherapy. *International Journal of Radiation Oncology*Biology*Physics* **72**, 996-1001, doi:http://dx.doi.org/10.1016/j.ijrobp.2008.02.047 (2008).
11. Shaffer, R. *et al.* Volumetric Modulated Arc Therapy and Conventional Intensity-modulated Radiotherapy for Simultaneous Maximal Intraprostatic Boost: a Planning Comparison Study. *Clinical Oncology* **21**, 401-407, doi:http://dx.doi.org/10.1016/j.clon.2009.01.014 (2009).
12. Shaffer, R. *et al.* A Comparison of Volumetric Modulated Arc Therapy and Conventional Intensity-Modulated Radiotherapy for Frontal and Temporal High-Grade Gliomas. *International Journal of Radiation Oncology*Biology*Physics* **76**, 1177-1184, doi:http://dx.doi.org/10.1016/j.ijrobp.2009.03.013 (2010).
13. Xing, S. M. C. a. X. W. a. C. T. a. M. W. a. L. Aperture modulated arc therapy. *Physics in Medicine & Biology* **48**, 1333 (2003).
14. Earl, M., Shepard, D., Naqvi, S., Li, X. & Yu, C. Inverse planning for intensity-modulated arc therapy using direct aperture optimization. *Physics in medicine and biology* **48**, 1075 (2003).
15. Daliang Cao and Muhammad, K. N. A. a. J. Y. a. F. C. a. D. M. S. A generalized inverse planning tool for volumetric-modulated arc therapy. *Physics in Medicine & Biology* **54**, 6725 (2009).

16 Marta, G. N. *et al.* Intensity-modulated radiation therapy for head and neck cancer: Systematic review and meta-analysis. *Radiotherapy and Oncology* **110**, 9-15, doi:https://doi.org/10.1016/j.radonc.2013.11.010 (2014).
17 Toledano, I. *et al.* Intensity-modulated radiotherapy in head and neck cancer: Results of the prospective study GORTEC 2004–03. *Radiotherapy and Oncology* **103**, 57-62, doi:https://doi.org/10.1016/j.radonc.2011.12.010 (2012).
18 Gupta, T. *et al.* Three-dimensional conformal radiotherapy (3D-CRT) versus intensity modulated radiation therapy (IMRT) in squamous cell carcinoma of the head and neck: A randomized controlled trial. *Radiotherapy and Oncology* **104**, 343-348, doi:https://doi.org/10.1016/j.radonc.2012.07.001 (2012).
19 Rathod, S. *et al.* Quality-of-life (QOL) outcomes in patients with head and neck squamous cell carcinoma (HNSCC) treated with intensity-modulated radiation therapy (IMRT) compared to three-dimensional conformal radiotherapy (3D-CRT): Evidence from a prospective randomized study. *Oral Oncology* **49**, 634-642, doi:https://doi.org/10.1016/j.oraloncology.2013.02.013 (2013).
20 Tribius, S. & Bergelt, C. Intensity-modulated radiotherapy versus conventional and 3D conformal radiotherapy in patients with head and neck cancer: Is there a worthwhile quality of life gain? *Cancer Treatment Reviews* **37**, 511-519, doi:https://doi.org/10.1016/j.ctrv.2011.01.004 (2011).
21 Paulino, A. C., Koshy, M., Howell, R., Schuster, D. & Davis, L. W. Comparison of CT- and FDG-PET-defined gross tumor volume in intensity-modulated radiotherapy for head-and-neck cancer. *International Journal of Radiation Oncology*Biology*Physics* **61**, 1385-1392, doi:https://doi.org/10.1016/j.ijrobp.2004.08.037 (2005).
22 Studer, G. *et al.* IMRT using simultaneously integrated boost (SIB) in head and neck cancer patients. *Radiation Oncology (London, England)* **1**, 7-7, doi:10.1186/1748-717X-1-7 (2006).
23 Wu, Q., Mohan, R., Morris, M., Lauve, A. & Schmidt-Ullrich, R. Simultaneous integrated boost intensity-modulated radiotherapy for locally advanced head-and-neck squamous cell carcinomas. I: dosimetric results. *International Journal of Radiation Oncology*Biology*Physics* **56**, 573-585, doi:https://doi.org/10.1016/S0360-3016(02)04617-5 (2003).
24 Mayo, C., Yorke, E. & Merchant, T. E. Radiation Associated Brainstem Injury. *International Journal of Radiation Oncology • Biology • Physics* **76**, S36-S41, doi:10.1016/j.ijrobp.2009.08.078.
25 Kirkpatrick, J. P., van der Kogel, A. J. & Schultheiss, T. E. Radiation Dose–Volume Effects in the Spinal Cord. *International Journal of Radiation Oncology • Biology • Physics* **76**, S42-S49, doi:10.1016/j.ijrobp.2009.04.095.
26 Deasy, J. O. *et al.* Radiotherapy Dose–Volume Effects on Salivary Gland Function. *International Journal of Radiation Oncology • Biology • Physics* **76**, S58-S63, doi:10.1016/j.ijrobp.2009.06.090.
27 Rancati, T. *et al.* Radiation Dose–Volume Effects in the Larynx and Pharynx. *International Journal of Radiation Oncology • Biology • Physics* **76**, S64-S69, doi:10.1016/j.ijrobp.2009.03.079.
28 Oelfke, U. & Bortfeld, T. Inverse planning for photon and proton beams. *Medical dosimetry* **26**, 113-124, doi:10.1016/s0958-3947(01)00057-7 (2001).

29  Craft, D. L., Hong, T. S., Shih, H. A. & Bortfeld, T. R. Improved Planning Time and Plan Quality Through Multicriteria Optimization for Intensity-Modulated Radiotherapy. *International Journal of Radiation Oncology\*Biology\*Physics* **82**, e83-e90, doi:https://doi.org/10.1016/j.ijrobp.2010.12.007 (2012).
30  Schreiner, L. J. On the quality assurance and verification of modern radiation therapy treatment. *Journal of medical physics/Association of Medical Physicists of India* **36**, 189 (2011).
31  Van Dye, J., Batista, J. & Bauman, G. S. Accuracy and uncertainty considerations in modern radiation oncology. *The Modern Technology of Radiation Oncology* **3**, 361-412 (2013).
32  Jensen, A. R., Nellemann, H. M. & Overgaard, J. Tumor progression in waiting time for radiotherapy in head and neck cancer. *Radiotherapy and Oncology* **84**, 5-10, doi:https://doi.org/10.1016/j.radonc.2007.04.001 (2007).
33  Bese, N. S., Hendry, J. & Jeremic, B. Effects of Prolongation of Overall Treatment Time Due To Unplanned Interruptions During Radiotherapy of Different Tumor Sites and Practical Methods for Compensation. *International Journal of Radiation Oncology • Biology • Physics* **68**, 654-661, doi:10.1016/j.ijrobp.2007.03.010.
34  González Ferreira, J. A., Jaén Olasolo, J., Azinovic, I. & Jeremic, B. Effect of radiotherapy delay in overall treatment time on local control and survival in head and neck cancer: Review of the literature. *Reports of Practical Oncology and Radiotherapy* **20**, 328-339, doi:10.1016/j.rpor.2015.05.010 (2015).
35  Fowler, J. F. & Lindstrom, M. J. Loss of local control with prolongation in radiotherapy. *International Journal of Radiation Oncology\*Biology\*Physics* **23**, 457-467, doi:https://doi.org/10.1016/0360-3016(92)90768-D (1992).
36  Fein, D. A. *et al.* Do overall treatment time, field size, and treatment energy influence local control of T1–T2 squamous cell carcinomas of the glottic larynx? *International Journal of Radiation Oncology\*Biology\*Physics* **34**, 823-831, doi:https://doi.org/10.1016/0360-3016(95)02205-8 (1996).
37  Ronneberger, O., Fischer, P. & Brox, T. U-net: Convolutional networks for biomedical image segmentation. *International Conference on Medical Image Computing and Computer-Assisted Intervention*, 234-241 (2015).
38  LeCun, Y. *et al.* Backpropagation applied to handwritten zip code recognition. *Neural computation* **1**, 541-551 (1989).
39  Long, J., Shelhamer, E. & Darrell, T. Fully convolutional networks for semantic segmentation. *Proceedings of the IEEE Conference on Computer Vision and Pattern Recognition*, 3431-3440 (2015).
40  Milletari, F., Navab, N. & Ahmadi, S.-A. V-net: Fully convolutional neural networks for volumetric medical image segmentation. *3D Vision (3DV), 2016 Fourth International Conference on*, 565-571 (2016).
41  He, K., Zhang, X., Ren, S. & Sun, J. Deep residual learning for image recognition. *Proceedings of the IEEE conference on computer vision and pattern recognition*, 770-778 (2016).
42  Huang, G., Liu, Z., van der Maaten, L. & Weinberger, K. Q. Densely Connected Convolutional Networks. *Proc Cvpr Ieee* **1**, 2261-2269, doi:10.1109/Cvpr.2017.243 (2017).

43    LeCun, Y., Bottou, L., Bengio, Y. & Haffner, P. Gradient-based learning applied to document recognition. *Proceedings of the IEEE* **86**, 2278-2324 (1998).
44    LeCun, Y., Haffner, P., Bottou, L. & Bengio, Y. in *Shape, contour and grouping in computer vision*    319-345 (Springer, 1999).
45    Nair, V. & Hinton, G. E. in *Proceedings of the 27th international conference on machine learning (ICML-10).*  807-814.
46    Nagi, J. *et al.* in *Signal and Image Processing Applications (ICSIPA), 2011 IEEE International Conference on.*  342-347 (IEEE).
47    Noh, H., Hong, S. & Han, B. Learning deconvolution network for semantic segmentation. *Proceedings of the IEEE International Conference on Computer Vision*, 1520-1528 (2015).
48    Prechelt, L. Automatic early stopping using cross validation: quantifying the criteria. *Neural Networks* **11**, 761-767 (1998).
49    Caruana, R., Lawrence, S. & Giles, C. L. Overfitting in neural nets: Backpropagation, conjugate gradient, and early stopping. *Advances in neural information processing systems*, 402-408 (2001).
50    Breiman, L. Bagging predictors. *Machine learning* **24**, 123-140 (1996).
51    Buja, A. & Stuetzle, W. Observations on bagging. *Statistica Sinica*, 323-351 (2006).
52    Van't Riet, A., Mak, A. C., Moerland, M. A., Elders, L. H. & van der Zee, W. A conformation number to quantify the degree of conformality in brachytherapy and external beam irradiation: application to the prostate. *International Journal of Radiation Oncology\* Biology\* Physics* **37**, 731-736 (1997).
53    Zhu, X. *et al.* A planning quality evaluation tool for prostate adaptive IMRT based on machine learning. *Medical physics* **38**, 719-726 (2011).
54    Appenzoller, L. M., Michalski, J. M., Thorstad, W. L., Mutic, S. & Moore, K. L. Predicting dose-volume histograms for organs-at-risk in IMRT planning. *Medical physics* **39**, 7446-7461 (2012).
55    Wu, B. *et al.* Improved robotic stereotactic body radiation therapy plan quality and planning efficacy for organ-confined prostate cancer utilizing overlap-volume histogram-driven planning methodology. *Radiotherapy and Oncology* **112**, 221-226 (2014).
56    Shiraishi, S., Tan, J., Olsen, L. A. & Moore, K. L. Knowledge-based prediction of plan quality metrics in intracranial stereotactic radiosurgery. *Medical physics* **42**, 908-917 (2015).
57    Moore, K. L., Brame, R. S., Low, D. A. & Mutic, S. Experience-Based Quality Control of Clinical Intensity-Modulated Radiotherapy Planning. *International Journal of Radiation Oncology\*Biology\*Physics* **81**, 545-551, doi:https://doi.org/10.1016/j.ijrobp.2010.11.030 (2011).
58    Shiraishi, S. & Moore, K. L. Knowledge-based prediction of three-dimensional dose distributions for external beam radiotherapy. *Medical physics* **43**, 378-387 (2016).
59    Wu, B. *et al.* Patient geometry-driven information retrieval for IMRT treatment plan quality control. *Medical Physics* **36**, 5497-5505, doi:10.1118/1.3253464 (2009).
60    Wu, B. *et al.* Data-Driven Approach to Generating Achievable Dose–Volume Histogram Objectives in Intensity-Modulated Radiotherapy Planning. *International Journal of Radiation Oncology\*Biology\*Physics* **79**, 1241-1247, doi:https://doi.org/10.1016/j.ijrobp.2010.05.026 (2011).

61    Wu, B. *et al.* Using overlap volume histogram and IMRT plan data to guide and automate VMAT planning: A head-and-neck case study. *Medical Physics* **40**, 021714-n/a, doi:10.1118/1.4788671 (2013).
62    Tran, A. *et al.* Predicting liver SBRT eligibility and plan quality for VMAT and 4π plans. *Radiation Oncology* **12**, 70, doi:10.1186/s13014-017-0806-z (2017).
63    Yuan, L. *et al.* Quantitative analysis of the factors which affect the interpatient organ-at-risk dose sparing variation in IMRT plans. *Medical Physics* **39**, 6868-6878, doi:10.1118/1.4757927 (2012).
64    Lian, J. *et al.* Modeling the dosimetry of organ-at-risk in head and neck IMRT planning: An intertechnique and interinstitutional study. *Medical Physics* **40**, 121704-n/a, doi:10.1118/1.4828788 (2013).
65    Folkerts, M. M., Gu, X., Lu, W., Radke, R. J. & Jiang, S. B. SU-G-TeP1-09: Modality-Specific Dose Gradient Modeling for Prostate IMRT Using Spherical Distance Maps of PTV and Isodose Contours. *Medical Physics* **43**, 3653-3654, doi:10.1118/1.4956999 (2016).
66    Folkerts, M. M. *et al.* Knowledge-Based Automatic Treatment Planning for Prostate IMRT Using 3-Dimensional Dose Prediction and Threshold-Based Optimization. *American Association of Physicists in Medicine* (2017).
67    Tol, J. P., Delaney, A. R., Dahele, M., Slotman, B. J. & Verbakel, W. F. Evaluation of a knowledge-based planning solution for head and neck cancer. *International Journal of Radiation Oncology• Biology• Physics* **91**, 612-620 (2015).
68    Krayenbuehl, J., Norton, I., Studer, G. & Guckenberger, M. Evaluation of an automated knowledge based treatment planning system for head and neck. *Radiation Oncology* **10**, 226 (2015).
69    Varian Medical Systems, I. RapidPlan Knowledge-Based Planning Frequently Asked Questions.  (2018). <https://www.varian.com/sites/default/files/resource_attachments/RapidPlanFAQs_RAD10321B.pdf>.
70    Li, X. *et al.* H-DenseUNet: Hybrid Densely Connected UNet for Liver and Tumor Segmentation from CT Volumes. *IEEE Transactions on Medical Imaging* (2018).
71    Long, T., Chen, M., Jiang, S. B. & Lu, W. Threshold-driven optimization for reference-based auto-planning. *Physics in medicine and biology* (2018).

**APPENDIX**

A.1. Details on deep learning architectures used in study

|  | HD U-net |  | Standard U-net |  | DenseNet |  |
| --- | --- | --- | --- | --- | --- | --- |
| Layer number | Layer type | Number features / channels | Layer type | Number features / channels | Layer type | Number features / channels |
|  |  |  |  |  |  |  |
| 1 | Input | 23 | Input | 23 | Input | 23 |
| 2 | Dense Conv | 39 | Conv | 32 | Dense Conv | 47 |
| 3 | Dense Conv | 55 | Conv | 32 | Dense Conv | 71 |
| 4 | Dense Downsample | 71 | Max Pooling | 32 | Dense Conv | 95 |
| 5 | Dense Conv | 87 | Conv | 64 | Dense Conv | 119 |
| 6 | Dense Conv | 103 | Conv | 64 | Dense Conv | 143 |
| 7 | Dense Downsample | 119 | Max Pooling | 64 | Conv | 72 |
| 8 | Dense Conv | 135 | Conv | 128 | Dense Conv | 96 |
| 9 | Dense Conv | 151 | Conv | 128 | Dense Conv | 120 |
| 10 | Dense Downsample | 167 | Max Pooling | 128 | Dense Conv | 144 |
| 11 | Dense Conv | 183 | Conv | 256 | Dense Conv | 168 |
| 12 | Dense Conv | 199 | Conv | 256 | Dense Conv | 192 |
| 13 | Dense Downsample | 215 | Max Pooling | 256 | Conv | 96 |
| 14 | Dense Conv | 231 | Conv | 512 | Dense Conv | 120 |
| 15 | Dense Conv | 247 | Conv | 512 | Dense Conv | 144 |
| 16 | Dense Conv | 263 | Conv | 512 | Dense Conv | 168 |
| 17 | Dense Conv | 279 | Conv | 512 | Dense Conv | 192 |
| 18 | U-net Upsample | 263 | U-net Upsample | 512 | Dense Conv | 216 |
| 19 | Dense Conv | 279 | Conv | 256 | Conv | 108 |
| 20 | Dense Conv | 295 | Conv | 256 | Dense Conv | 132 |
| 21 | U-net Upsample | 215 | U-net Upsample | 256 | Dense Conv | 156 |
| 22 | Dense Conv | 231 | Conv | 128 | Dense Conv | 180 |
| 23 | Dense Conv | 247 | Conv | 128 | Dense Conv | 204 |
| 24 | U-net Upsample | 167 | U-net Upsample | 128 | Dense Conv | 228 |
| 25 | Dense Conv | 183 | Conv | 64 | Conv | 114 |
| 26 | Dense Conv | 199 | Conv | 64 | Dense Conv | 138 |
| 27 | U-net Upsample | 119 | U-net Upsample | 64 | Dense Conv | 162 |
| 28 | Dense Conv | 135 | Conv | 32 | Dense Conv | 186 |
| 29 | Dense Conv | 151 | Conv | 32 | Dense Conv | 210 |
| 30 | Conv | 1 | Conv | 1 | Dense Conv | 234 |
| 31 | | | | | Conv | 117 |
| 32 | | | | | Dense Conv | 141 |
| 33 | | | | | Dense Conv | 165 |
| 34 | | | | | Dense Conv | 189 |
| 35 | | | | | Dense Conv | 213 |
| 36 | | | | | Dense Conv | 237 |
| 37 | | | | | Conv | 119 |
| 38 | | | | | Dense Conv | 143 |
| 39 | | | | | Dense Conv | 167 |
| 40 | | | | | Dense Conv | 191 |
| 41 | | | | | Dense Conv | 215 |
| 42 | | | | | Dense Conv | 239 |
| 43 | | | | | Conv | 120 |
| 44 | | | | | Conv | 1 |

Table 7: Details of deep learning architectures. Dense Conv and U-net Upsample follow the notation outlined in Figure 1. All convolutions mentioned in this table use 3 x 3 x 3 kernels and are followed by the ReLU non-linear activation.